\documentclass[11pt]{article}
\usepackage{graphicx}

\begin{document}

\def\bfej{\mbox{\boldmath$\varepsilon$}}
\newcommand{\ccbar}{c\bar c}
\newcommand {\sla}[1]{ #1 \!\!\!/}
\newcommand{\jgmm}{\jp\to \gamm MM}
\newcommand{\jp}{J/\psi}
\newcommand{\st}[1]{|#1\rangle}
\newcommand{\jgpp}{\jp\to\gamma\phi\phi}
\def \qqbar {q\bar q}
\newcommand{\str}[1]{|#1\rangle}
\newcommand{\elmnt}[3]{\langle #1|#2|#3\rangle}
\newcommand{\cp}{CP~}
\newcommand{\cpv}{CP violation~}

\vskip 10mm
\begin{center}
{\bf\Large CP  Test in $\jp\to \gamma \phi\phi$ Decay  } \\
\vskip 10mm
J.P. Ma, R.G. Ping and B.S. Zou \\
\vskip10pt
{\small{\it CCAST(World Lab.), P.O.Box 8730, Beijing 100080, China;\\
Institute of High Energy Physics, Academia
Sinica, Beijing 100039, China; \\
Institute of Theoretical Physics , Academia
Sinica, Beijing 100080, China \ \ \ }} \\
\end{center}

\vskip 0.4 cm

\begin{abstract}
We propose to test CP symmetry in the decay $\jp\to \gamma \phi\phi$,
for which large data sample exists at BESII, and a data sample
of $10^{10}$ $J/\psi$'s will be collected with BESIII and CLEO-C
program. We suggest some CP asymmetries in this decay mode
for CP test. Assuming that CP violation is introduced
by the electric- and chromo-dipole moment of charm quark,
these CP asymmetries can be predicted by using
valence quark models. Our work shows a possible
way to get information about the electric- and chromo-dipole moment of charm quark,
which is little known.
Our results show that
with the current data sample of $J/\psi$,
electric- and chromo-dipole moment can be probed
at order of $10^{-13}e~cm$. In the near future
with a $10^{10}$ data sample, these moments can be
probed at order of $10^{-14}e~cm$.
\end{abstract}
\par\vfil\eject

\par
CP violation was first discovered in $K$-system in the last century\cite{kaon}.
At the beginning of this century, CP violation was also found in $B$-system
thanks to two $B$-factories\cite{BCP}. Except these, there is no experimental
indication for CP violation. With our current knowledge interactions
between elementary particles are described by the Standard Model(SM).
In SM CP symmetry is violated by complex CKM matrix elements.
Although SM is successful and can explain CP violation
effects found in $K$- and $B$-system, it is generally believed
that SM is only an effective theory for energy scale around
$100$GeV. It is expected that new physics beyond SM
will emerge. Therefore, it is important to find CP violation somewhere else,
where SM predicts no or tiny CP violation, because this can be
an indication for new physics.
\par
It is expected that effect of CP violation is very small in general.
Hence, Experimental study of CP violation needs large data samples.
With progress at BEPC and CLEO-C large data samples of
$J/\psi$ decays exist or will be collected. At BEPC with BESII
detector there are already $5.8\times 10^7$ $J/\psi$-decay events.
With upgraded BEPC\cite{BES3} and CLEO-C program\cite{CLEOC},
a data sample of $10^9\sim 10^{10}$ $J/\psi$ will be collected. These huge data samples
are very suited for CP test. But, not every decay mode
of $J/\psi$ can be used for CP test. For $J/\psi$ decay
into a particle and its antiparticle, CP test is not possible
if these particles are spinless or their polarizations
are not observed\cite{BLMN}. CP test is only possible
if polarizations of decay products are measured.
The decay $J/\psi\to \Lambda\bar\Lambda$ is an example, where
the polarizations can be measured though subsequential decays of $\Lambda$
and $\bar\Lambda$.  CP test
with this decay mode has been studied in \cite{HeMa}.
CP test can be made for three-body decays, even without knowing
polarizations of decay products. In this letter we propose
to study CP test in the decay $J/\psi\to\gamma\phi\phi$.
This decay can provide useful information about
electric dipole moment (EDM) of charm quark.
The reason for choosing this channel is because $\phi$ is a very narrow
resonance just above $K\bar K$ threshold and can be clearly identified
by its $K^+K^-$
decay mode in experiment. In principle $J/\psi\to\gamma\rho\rho$
could also serve the purpose, but experimentally the broad width of $\rho$ meson
makes it impossible to get clean sample for the channel.
By using large charmonia  data samples it has been
also discussed in \cite{Oth} about possible CP test, e.g., $\psi'\to J/\psi\pi\pi$,
and in
\cite{CGR} about parity test in the decay mode $J/\psi\to\gamma\ell^+\ell^-$.
\par
From SM EDM of quarks and leptons are very small(See Reviews in \cite{HeMc,BernSu}, and
references therein). If EDM
of quark is found to be nonzero, it is likely an indication
for new physics. Because the operator for EDM does not
converse helicities of quarks, its effect
is suppressed in a high energy process
by a factor $m_q/E$, where $m_q$ is the quark mass and $E$
is a large energy scale. For light quarks, useful information
can be obtained through measurement of EDM of neutron
\cite{HeMc}.
So far there is no experimental information
about EDM's of heavy quarks, like charm- and bottom quark.
$J/\psi$ decays can provide information of EDM of charm quark
and has the advantage that the effect of EDM will be not suppressed,
because the large energy scale is around $m_c$. Since in the decay
a $c\bar c$ pair is annihilated into a photon and gluons, it also
provides a way to detecting chromodipole moment of charm quark.
In this letter we assume that CP violation is introduced
by electric- and chromo-dipole moment and study
their effect in the decay. The moments are defined by effective
Lagrangian:
\begin{eqnarray}
L_{CP} = -i\frac{d_c}{2}\bar c\gamma_5\sigma_{\mu\nu}F^{\mu\nu}c
          -i\frac{\tilde{d}_c}{2} \bar
c\gamma_5\sigma_{\mu\nu}G^{\mu\nu}c,
\end{eqnarray}
where $d_c$ is the electric dipole moment, $\tilde d_c$ is the chromodipole
moment. With the decay mode, it is also possible to detect
the electric dipole moment of $s$-quark. But its effect
is suppressed by $m_s/m_c$. In this letter
we will only consider dipole moments of $c$-quark.
\par
Unfortunately, predictions for the decay $J/\psi\to \gamma\phi\phi$ can not be
made without models, especially, because there is no reliable
method in theory how to determine the formation of bound states like $\phi$.
For $J/\psi$ a good approximation is to use a nonrelativistic wave-function
for describing  $J/\psi$ as a bound state of a $c\bar c$ pair, because
$c$-quark can be taken as a heavy quark. In this work we will also
take this approximation for $\phi$ meson by taking it as a nonrelativistic
bound state of a $s\bar s$ quark pair, where the $s$-quark should be taken
as a constituent quark, but not as a current quark.
The constituent $s$-quark mass should be taken as $m_s \approx m_\phi/2$.
This approximation
was employed in radiative decay into a hadron\cite{KN}. In this approximation,
the decay amplitude is factorized into a partonic part which is for the
process $c\bar c\to \gamma s\bar s s\bar s$, and a nonperturbative part
which is related to wave functions of hadrons at the origin. The partonic
part can be calculated with perturbative theory. The wave functions
at the origin can be determined by experiment. In this letter we
will take this approximation. It should be noted that with this approximation
reliable predictions for various distributions may not be obtained.
But one may expect that for our integrated observables defined in below
it can be a good approximation. Especially, our integrated
observables will not depend on wave functions at the origin.
\par
We consider the decay in the rest-frame of $J/\psi$
\begin{equation}
J/\psi(P) \to \gamma(k) +\phi(p_1) +\phi(p_2),
\end{equation}
where momenta are given in brackets. Because
the two $\phi$ mesons are identical particles,
we require $p_1^0 > p_2^0$ to distinguish them in experiment.
The decay amplitude can be written as:
\begin{equation}
 {\cal T}(J/\psi\to\gamma\phi\phi) = A_i({\bf p_1},{\bf p_2}) \varepsilon_i,
\end{equation}
where $\varepsilon_i(i=1,2,3)$ is the spacial component of the polarization
vector of $J/\psi$. With the amplitude, a density matrix for the
decay can be defined as:
\begin{equation}
   R_{ij} ({\bf p_1},{\bf p_2}) =\sum_{spin}  A_i({\bf p_1},{\bf p_2})
                                 A_j^* ({\bf p_1},{\bf p_2}).
\end{equation}
The sum is over spins of all particles in the final state.
This matrix can
be decomposed in terms of tensor built with ${\bf p_1}$, ${\bf p_2}$,
$\delta_{ij}$ and $\varepsilon_{ijk}$. A similar decomposition
can be found in \cite{BLMN}.
The density matrix contains all information about the decay.
Any experimental observable can be predicted if we know the density
matrix for production. We consider the situation where $J/\psi$ is
produced through $e^+e^-$ annihilation in their center-mass frame,
as at BEPC or CLEO-C. The density matrix $\rho_{ij}$ for the production takes
the form:
\begin{equation}\label{}
\rho_{ij}={1\over 3}\delta_{ij}-{1\over 2}(\hat{k}_{+i}\hat
k_{+j}-{1\over 3}\delta_{ij}),
\end{equation}
where $\hat{k}_+$ denotes the moving direction of $e^+$. With
these density matrices, the expectation value of
any experimental observable $O$
constructed with ${\bf k_+}$, ${\bf p_1}$ and ${\bf p_2}$
can be predicted by
\begin{equation}\label{obs}
\langle O\rangle =  {1\over \cal{N}}\int d\Gamma (2\pi)^4\delta(P-p_1-p_2-k)
  O \rho_{ij}R_{ji},
\end{equation}
where $d\Gamma$ is the differential phase-space of $\gamma\phi\phi$
with $p_1^0 > p_2^0$, ${\cal N}$ is a normalization factor so
that $ \langle 1 \rangle =1$.
\par
Symmetries will constrain the form
of the density matrix. If CP symmetry holds, we have
\begin{equation}
 R_{ij} ({\bf p_1},{\bf p_2}) =R_{ij} (-{\bf p_1},-{\bf p_2}).
\end{equation}
For CP test one can check this relation to see if it holds.
A convenient way to check this is to use some integrated observables.
We find two \cp odd observables which we are particularly
interested in:
\begin{equation}\label{}
O_1={\bf \hat k_+}\cdot {\bf \hat p_1} {\bf \hat k_+}\cdot
  ({\bf\hat p_1}\times {\bf\hat p_2}),
~~O_2={\bf \hat k_+}\cdot {\bf \hat p_2} {\bf \hat k_+}\cdot
  ({\bf\hat p_1}\times {\bf\hat p_2}),
\end{equation}
where ${\bf\hat p_1}$ or ${\bf\hat p_2}$ are the direction
of ${\bf p_1}$ or ${\bf p_2}$ respectively.
From these oberservables, One can define \cp-asymmetry as
\begin{equation}\label{cpa}
B_i=\langle \theta(O_i)-\theta(-O_i)\rangle~~(i=1,2),
\end{equation}
where $\theta (x)=1$ if $x>0$ and is zero if $x<0$.
If these asymmetries are not zero, CP symmetry is violated.
In this letter, we will make predictions for these asymmetries
by taking CP violating interactions as given in Eq.(1).
\par

\begin{figure}[hbt]
\centering
\includegraphics[width=6cm]{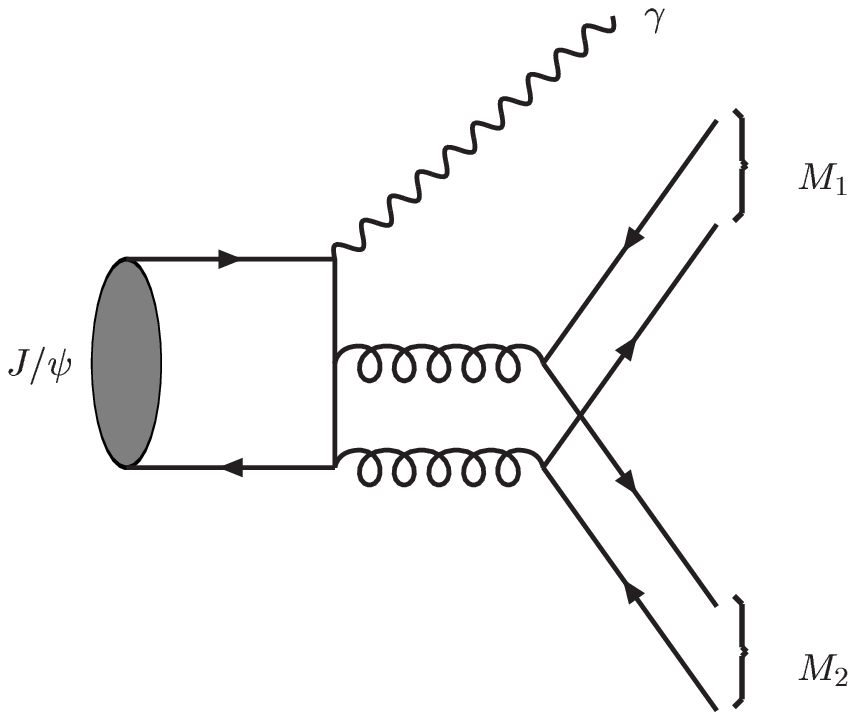}
\caption{A typical Feynman diagram for $J/\psi\to\gamma\phi\phi$,
other diagrams are obtained by changing attachments of gluons to
different places. Replacing $c\bar c g$ and $c\bar c \gamma$ coupling
with the corresponding dipole interaction one obtains CP violating
amplitude.}
\label{Feynman-dg1}
\end{figure}
\par
\par
\vskip20pt
As discussed before, we will use nonrelativistic wave-functions
for bound states. A typical Feynman diagram is given in Fig.1.
From which one can read out the contribution to the scattering  amplitude
for $c(q_1)+\bar c(q_2) \to \gamma(k) +s(k_1) +\bar s(k_2) + s(k_3)
   +\bar s(k_4)$, which takes the form:
\begin{equation}
{\cal T}_1(c(q_1)+\bar c(q_2) \to \gamma(k) +s(k_1) +\bar s(k_2) + s(k_3)
   +\bar s(k_4) ) = \bar v(q_2) C^{ab,\mu\nu} u(q_1)
                       \bar u(k_1) \gamma_\mu T^a v(k_4)
                       \bar u(k_3) \gamma_\nu T^b v(k_2),
\end{equation}
where $C^{ab,\mu\nu}$ is a matrix with Dirac- and color indices.
To obtain the contribution from Fig.1 to the hadronic decay amplitude,
one needs to project the quark pairs into hadron states.
The projection is made by the replacement\cite{GKPR}:
\begin{eqnarray}
 u_i(q_1) \bar v_j(q_2) &\to & \frac {1}{3 (2m_c)^{3/2}}
 \psi_{J/\psi} (0) \left [ (\gamma\cdot q_1 + m_c)\gamma\cdot \varepsilon(P)
                   (-\gamma\cdot q_2 +m_c) \right ]_{ij},
\nonumber\\
v_i(k_2) \bar u_j(k_1) &\to & \frac {1}{3 (2m_s)^{3/2}}
 \psi^*_{\phi} (0) \left [ (-\gamma\cdot k_2 + m_s)\gamma\cdot \varepsilon^*(p_1)
                   (\gamma\cdot k_1 +m_s) \right ]_{ij},
\nonumber\\
v_i(k_4) \bar u_j(k_3) &\to & \frac {1}{3 (2m_s)^{3/2}}
 \psi^*_{\phi} (0) \left [ (-\gamma\cdot k_4 + m_s)\gamma\cdot \varepsilon^*(p_2)
                   (\gamma\cdot k_3 +m_s) \right ]_{ij},
\end{eqnarray}
where $ij$ stand for Dirac- and color indices, $\varepsilon(P)$,
$\varepsilon(p_1)$ and $\varepsilon(p_2)$ is
the polarization vector of $J/\psi(P)$, $\phi(p_1)$ and $\phi(p_2)$,
respectively. $\psi_{J/\psi}$($\psi_\phi$) is the wave function
of $J/\Psi$($\phi$).
In the projection momenta of quarks and masses should be taken as:
\begin{eqnarray}
q_1 &=& q_2 = \frac{P}{2},\ \ \  m_c =\frac{M_{J/\psi}}{2},
\nonumber\\
k_1 &=& k_2 = \frac{p_1}{2}, \ \ \ m_s =\frac{M_\phi}{2},
\nonumber\\
k_3 &=& k_4 = \frac{p_2}{2}.
\end{eqnarray}
It should be noted that from the amplitude in Eq.(10)
there is another contribution to the hadronic amplitude which
is obtained by exchanging $p_1\leftrightarrow p_2$.
The calculation is straightforward.
The total decay amplitude
is too complicated to be given here. A numerical program
for the amplitude can be obtained from authors.
\par
With the approximation explained in detail, numerical results
can be obtained. We use $\alpha_s(m_c) =0.28$ and masses
which are fixed by experimental values of hadrons given
in Eq.(12). We obtain the decay width:
\begin{equation}
\Gamma(J/\psi\to\gamma\phi\phi) = 1.83\times 10^{-3}
      \vert \psi_{J/\psi}(0)\vert^2 \vert \psi_{\phi}(0)\vert^4{\rm GeV}^{-8}.
\end{equation}
If we use the value of $\vert \psi_{J/\psi}(0)\vert^2=0.055{\rm GeV}^3$,
determined from study with potential models\cite{EQ} and lattice QCD\cite{BSK},
and experimental results of $\Gamma_{tot}=87{\rm KeV}$ and
$Br(J/\psi\to\gamma\phi\phi) =4.0\times 10^{-4}$, we can obtain
\begin{equation}
\vert \psi_{\phi}(0)\vert^2 \approx 0.019{\rm GeV}^3.
\end{equation}
It should be noted
that in our approximation the CP asymmetries in Eq.(9)
do not depend on these wave functions at the origin. With
the same parameters we obtain:
\begin{eqnarray}
B_1 &=& 4.2\left [ \frac { d_c}{10^{-10} e~cm}\right ] -1.2
  \left [ \frac{ \tilde d_c}{10^{-10} e~cm}\right ] ,
\nonumber\\
B_2 &=& -3.9 \left [ \frac { d_c}{10^{-10} e~cm}\right ]
    +1.3 \left [ \frac{ \tilde d_c}{10^{-10} e~cm}\right ].
\end{eqnarray}
A statistic sensitivity to $d_c$ and $\tilde d_c$ can be determined
from these results by requiring that the asymmetry generated by these
dipole moments should be larger than the statistical error. The statistical
error for these asymmetries is given by
\begin{equation}
\delta B_i =\sqrt{\frac{1}{N_{events}}},
\end{equation}
where $N_{events}$ is the number of the decay events.
With the $5.8\times 10^7~J/\psi$ data sample at BES, the sensitivities of
our CP asymmetries to these dipole moments are
\begin{equation}
d_c \sim 1.4\times 10^{-13} e~cm, \ \ \ \
\tilde d_c \sim 4.5\times 10^{-13} e~cm.
\end{equation}
With a $10^{10}$ data sample which will be collected in the near future,
the sensitivities are:
\begin{equation}
d_c \sim 1.2\times 10^{-14} e~cm, \ \ \ \
\tilde d_c \sim 3.6 \times 10^{-14} e~cm.
\end{equation}
\par
It is instructive to compare these sensitivities to the upper
bound of EDM of neutron\cite{nedm} and electron\cite{eedm} from
experiment. The upper bounds read:
\begin{eqnarray}
d_n & <& 6.3\times 10^{-26} e~cm,
\nonumber\\
d_e &<& 4.3\times 10^{-27} e~cm.
\end{eqnarray}
The implication of these upper bounds for physics beyond SM
has been studied extensively(See e.g., \cite{AKL} and references
therein). Predictions from any reasonable extension of SM
for EDM of light quarks and electron should not
be larger than these upper bounds. Comparing these upper bounds,
the sensitivities are too large. However, in some extensions
of SM predicted EDM of  a quark is proportional to
the mass of the quark(e.g., see \cite{BZ,CKP}). It is possible
to obtain EDM of $c$-quark which is comparable to these sensitivities.
Hence, CP test proposed in this letter will provide
information for possible new physics beyond SM.
\par
To summarize: In this letter we have proposed to test CP symmetry
in the decay $J/\psi\to\gamma\phi\phi$ by using large data samples
of $J/\psi$ which is already collected at BES and will be
collected
with BESIII and CLEO-C program. We introduce
some CP asymmetries for CP test with this decay mode.
Assuming that CP violation is introduced
by the electric- and chromo-dipole moment of charm quark, we
give predictions for these asymmetries by using wave functions
for hadrons. Experimental information is available
for electric dipole moments of light quarks through measurement
of the electric dipole moment of neutron, but there is no
information for electric dipole moments of heavy quarks.
Our work shows a possible way to obtain this information.
With the current data sample of $J/\psi$ at BES, the
electric- and chromo-dipole moment can be probed
at order of $10^{-13}e~cm$. In the near future
with a $10^{10}$ data sample, these moments can be
probed at order of $10^{-14}e~cm$.
\par\vskip20pt
{\bf Acknowledgements}
\par
One of us(J.P. Ma) would like to thank Prof. X.G. He and O. Nachtmann
for useful discussions.
This work is supported by National Nature
Science Foundation of P. R. China.

\par\vfil\eject

\end{document}